\documentclass[letterpaper]{article}

\usepackage[T1]{fontenc}

\usepackage{geometry}
\usepackage{setspace}
\usepackage[normalem]{ulem}

\usepackage[articletitle=true]{achemso}

\usepackage{graphicx}
\usepackage{float}
\newfloat{scheme}{htbp}{los}
\floatname{scheme}{Scheme}
\floatname{chart}{Chart}
\newfloat{graph}{htbp}{loh}

\usepackage{chemformula} % Formulas using \ch{}

\usepackage{amsmath,amssymb,amsfonts}%

\usepackage[version = 4]{mhchem} % Formulas using \ce{}

\let\polishl\l
\renewcommand{\l}{\lambda}

\newcommand{\be}{\begin{equation}}
\newcommand{\ee}{\end{equation}}
\newcommand{\bea}{\begin{eqnarray}}
\newcommand{\eea}{\end{eqnarray}}

\def\G{\Gamma}
\def\d{\delta}

\def\e{\epsilon}

\def\l{\lambda}

\def\r{\rho}

\def\W{\Omega}

\def\blk{{\mathbf k}}

\def\blQ{{\mathbf Q}}

\def\1op{\hat{\mathbbm{1}}}

\def\AA{\mathring{\mathrm{A}}}

\def\1op{\hat{\mathbbm{1}}}

\def\bz{\mathbf 0}

\usepackage{authblk}
\author[1,2]{Carlos Betancur}
\author[1,2]{Gianluca Stefanucci}
\author[1,2]{Enrico Perfetto*}

\affil[1]{Dipartimento di Fisica, Universit{\`a} di
Roma Tor Vergata, Via della Ricerca Scientifica 1,
00133 Rome, Italy}

\affil[2]{INFN, Sezione di Roma Tor Vergata, Via della Ricerca Scientifica
1, 00133 Rome, Italy}

\title{Excitonic effects in the photocarriers dynamics 
\\
of two-dimensional materials}
\date{*Email: enrico.perfetto@roma2.infn.it}

\begin{document}

\maketitle

\begin{abstract}
We investigate the role of excitonic correlations in shaping 
the ultrafast dynamics of photoexcited carriers in semiconductors.
Conventional approaches describe relaxation within single-particle 
frameworks, where electron–electron and electron–phonon scattering
drive thermalization toward Fermi–Dirac distributions, neglecting 
electron–hole correlations that dominate near band edges. 
We introduce a two-particle framework based on excitonic Bloch equations (XBE)
that captures carrier-phonon scattering and explicitly accounts for exciton formation.
Applying this approach to non-resonantly photoexcited WSe$_{2}$
monolayers, we reveal qualitatively different carrier relaxation 
pathways: in contrast to state-of-the-art 
methods, XBE predict enhanced intervalley scattering and dominant
carrier population in Q valleys over K valleys, in  agreement 
with time-resolved ARPES experiments.
Moreover, the  momentum-distribution of thermalized carriers is 
shaped by exciton wavefunctions
rather than  by Fermi–Dirac statistics, signaling 
the formation of a correlated non-equilibrium state.
These results establish excitonic
correlations as a key mechanism governing photocarrier dynamics
in excitonic materials.
\end{abstract}

 \section*{Keywords}

 \
 Two-dimensional semiconductors, ultrafast carrier dynamics, exciton 
 dynamics, exciton-phonon interaction 

% \section*{Abbreviations}
% 
% Some journals require a list of abbreviations: these normally should be given
% immediately after the keyswords (if required).

\section{Introduction}

Unraveling carrier dynamics following photoexcitation in 
semiconductors remains a central challenge in modern condensed-matter and materials
science~\cite{Caruso_2026}. These nonequilibrium processes determine how absorbed light
is converted into charge, heat, and chemical energy, ultimately 
setting the fundamental performance limits of optoelectronic,
photovoltaic, and photocatalytic 
technologies~\cite{ponseca,D1CS01164B,C9CS00254E,joule,annurev103742}.

The ultrafast motion of photogenerated electrons and holes is 
intrinsically complex because these quasiparticles are not 
independent: they interact both among themselves and with 
nuclear degrees of freedom. As a result, a rich variety 
of processes emerge on comparable timescales and compete during 
relaxation~\cite{HaugKochbook,Koch2006,KIRA2006155}. 
These include hot-carrier cooling toward the band 
edges via phonon emission and electron-hole (e-h) pair 
generation~\cite{cooling1,cooling2,nie2014}, 
the buildup of dynamical Coulomb 
screening~\cite{screening3,PhysRevLett.128.016801,PhysRevB.111.045147}, 
quasiparticle dressing~\cite{dressing2,dressing1},
exciton 
formation~\cite{Steinleitner,C6NR02516A,cha,D0CP03220D,trovatello2020,Wallauer2,GOSETTI2025100777}, nuclear 
motion~\cite{ishioka2009coherent,trovatello,cohexp5}, and lattice 
heating~\cite{heating2,PhysRevB.103.064305}.

The theoretical description of such nonequilibrium dynamics is
highly challenging, as multiple interactions---light--matter,
electron--electron (e--e), electron--phonon (e--p), and phonon--phonon ---must
be treated simultaneously in a nonperturbative and controlled 
framework, while remaining computationally tractable. 
In this context a quantitatively accurate description
of the energy exchange between carriers and lattice vibrations becomes
particularly critical,
as it governs decoherence, dissipation, and the establishment
of detailed balance during relaxation.

The nonequilibrium Green’s function (NEGF) 
formalism~\cite{svl-book}
has emerged as a powerful first-principles framework for simulating coupled
electron–phonon dynamics. NEGF is based on the Kadanoff–Baym 
equations (KBE)~\cite{kadanoff1962quantum}, which govern the temporal evolution of electronic and 
phononic Green’s functions~\cite{PhysRevX.13.031026}. However, the unfavorable cubic
scaling of KBE with propagation time prevents direct ab initio applications
to realistic materials. Practical approaches therefore rely on controlled
approximations to simplify the KBE. This leads to a hierarchy of methods ranging from non-Markovian
schemes — such as density-matrix equations based on the generalized Kadanoff–Baym
ansatz~\cite{PhysRevB.105.125134,PhysRevB.105.125135,perfetto_2023} — to Markovian descriptions, including 
semiconductor electron–phonon equations (SEPE)~\cite{sepe}, semiconductor Bloch 
equations 
(SBE)~\cite{HaugKochbook,PhysRevB.52.5624,PhysRevB.84.205406,Marini2013,Sangalli_2015,mocatti2025nonequilibrium}, and
Boltzmann equations 
(BE)~\cite{PhysRevB.102.024308,PhysRevResearch.3.023072,perturbo,caruso,Caruso31122022}.
With current computational capabilities, Markovian treatments 
remain the only viable route for ab initio simulations of
crystalline systems described by Brillouin-zone samplings involving
thousands of $\mathbf{k}$-points.

A common feature of  Markovian approaches 
is that e–e and e–p scattering processes are treated within 
 GW~\cite{PhysRevLett.128.016801} and
Fan--Migdal~\cite{PhysRevLett.127.036402}
approximations.
Here inelastic scattering drives relaxation toward a quasi-steady state 
in which carriers and phonons are distributed according to 
non-interacting Fermi--Dirac
and Bose--Einstein statistics~\cite{sepe}. Paradoxically, this long-time limit 
corresponds to an effectively uncorrelated state, despite the fact
that correlations are responsible for the relaxation process itself.
This limitation becomes particularly severe in excitonic materials.
As electrons and holes approach the band edges they lose their free-carrier character 
as they become bound into excitonic states~\cite{PhysRevB.65.035303,PhysRevB.92.235208,brem}. 
In this regime, excitonic correlations qualitatively reshape the 
carrier distribution, driving it away from a Fermi–Dirac form.
At low excitation densities and temperatures,
the distribution of relaxed conduction electrons follows the
shape of the lowest-energy 
exciton wavefunction $A_{cv\mathbf{k}}$ (see also Supporting Note~1), namely
\begin{equation}
f_{c \mathbf{k}} \approx N \sum_{v} \left| A_{cv\,\mathbf{k}} 
\right|^{2},
\end{equation}
where $N$ is the population of the lowest excitonic state
and $c$ ($v$) label conduction (valence)
bands. 
Capturing this physics requires a treatment of e–e interactions beyond the
GW approximation.

   \begin{figure}[tbp]
   \centering
   \includegraphics[width=0.99\textwidth]{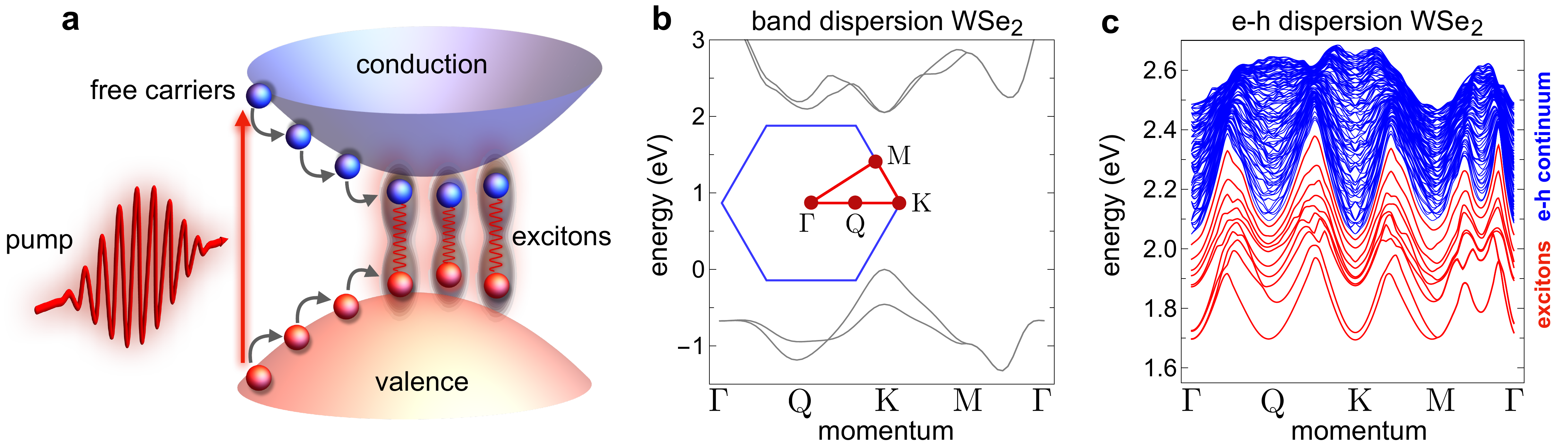}
   \caption{
   Panel a: Schematic of carrier dynamics following above-bandgap photoexcitation.
   Photoexcited electrons and holes initially relax by dissipating 
   excess energy through phonon emission; as they 
   approach the band edges, they further reduce their energy via the formation of
   bound excitons.
   Panel b: Electronic band dispersion of the two lowest conduction
   bands and the two highest valence bands of monolayer WSe$_{2}$ along
   the path in the first Brillouin zone indicated in the inset. 
   Band energies are computed as described in the Supporting Note~4. 
   Panel c:  Electron–hole excitation spectrum obtained from the solution
   of the Bethe–Salpeter equation [Eq.~(\ref{eigeneqX})] along the same path 
   as in panel b. The spectrum comprises bound excitons (red curves) and
   quasi-free electron–hole states (blue curves). 
   The 10 lowest excitonic branches and the 200 lowest quasi-free electron–hole
   branches are shown.
   }
   \label{fig1}
   \end{figure}

In this work we investigate for the first time the impact of excitonic 
correlations on the carrier dynamics of photoexcited semiconductors.
Our approach builds on the recently developed excitonic Bloch
equations (XBE)~\cite{10.21468/SciPostPhys.18.1.009}, which describe the time evolution of e-h 
populations within a correlated two-particle framework.
By projecting these populations onto the 
one-particle sector, we reconstruct the transient momentum-space distribution
of electrons and holes during and after photoexcitation. 
Excitonic correlations are incorporated through the inclusion of 
T-matrix scattering processes, which capture multiple e-h
interactions and are therefore particularly suited to 
describing bound excitons~\cite{Steinleitner,C6NR02516A,cha,D0CP03220D,trovatello2020}.

As a prototypical test case, we focus on a monolayer of WSe$_2$,
a paradigmatic excitonic material hosting a rich manifold of
strongly bound exciton states. To assess the capabilities of XBE,
we compare the resulting carrier dynamics with those obtained from 
SEPE~\cite{sepe}, from which the SBE and BE follow under additional approximations.
We focus on weak above-bandgap photoexcitation generating
a low excitation density. This textbook regime is traditionally
considered well-understood: GW correlations (responsible for 
excited-state screening and e-e scattering 
processes~\cite{PhysRevLett.128.016801,perfetto_2023,mocatti2025nonequilibrium})
play a negligible role, while hot-carrier relaxation
toward the band edges is governed
by electron–phonon interactions.
We find that XBE and SEPE yield qualitatively 
different dynamics, not only at long times—where SEPE predicts
thermalization towards a Fermi–Dirac distribution whereas XBE predict 
exciton wavefunction-like distribution—but already
on sub-picosecond timescales. Immediately
after photoexcitation, XBE points to highly efficient intervalley
scattering, resulting in a larger population in the Q valleys 
than in the K valleys. Remarkably, this finding is consistent 
with recent time-resolved ARPES measurements~\cite{madeo2020}, but in sharp 
contrast to SEPE predictions, which instead favor K-valley occupation.

Our results demonstrate that excitonic correlations play 
a decisive role in shaping ultrafast carrier dynamics, even in
regimes traditionally considered dominated by electron–phonon 
scattering. More broadly, they establish the need for explicitly excitonic
frameworks to achieve a predictive description of carrier dynamics
in semiconductors and, in particular, in low-dimensional materials
hosting strongly bound excitons.

\section{Results and discussion}

To incorporate excitonic effects into carrier dynamics, we adopt
a two-particle Green’s-function framework that describes the photoinduced
evolution of the occupations
of all e–h states, encompassing both quasi-free 
e–h pairs
and bound excitons.
A suitable basis for these states is obtained from the solution of
the finite-momentum Bethe–Salpeter equation (BSE)
\begin{equation}
(\e_{c\blk+\blQ}-\e_{v\blk})A^{\l\blQ}_{cv\blk}-
\sum_{c'v'\blk'}
K^{\rm HSEX,\blQ}_{cv\blk,c'v'\blk'}A^{\l\blQ}_{c'v'\blk'}=
E_{\l\blQ}A^{\l\blQ}_{cv\blk},
\label{eigeneqX}
\end{equation}
where $\e_{v(c)\blk}$ is the valence (conduction) band dispersion, 
$\blQ$ is the  e-h center-of-mass momentum, 
$A^{\l\blQ}$ denotes the e-h wavefunction with energy $E_{\l\blQ}$, and 
$K^{\rm HSEX}$ is the Hartree plus screened-exchange (HSEX) kernel.
The discrete sector of the BSE spectrum describes bound 
excitonic states, while the continuum 
corresponds to unbound, quasi–free e-h pairs.
The XBE~\cite{10.21468/SciPostPhys.18.1.009} provide an efficient Markovian,
first-principles approach to compute
the time-dependent occupations
$N_{\l \blQ}(t)$ of all the eigenstates 
of Eq.~(\ref{eigeneqX}) under the action of an external
field and phonon-mediated scattering (see below).
As shown in the Supporting Note~1, the corresponding
single-particle carrier distributions follow as
\be
f_{c \mathbf{k}}(t) =\sum_{\l \blQ v}N_{\l 
\blQ}(t)|A^{\l\blQ}_{cv\blk-\blQ}|^{2} \quad, \quad f_{v 
\mathbf{k}}(t) =1-\sum_{\l \blQ c}N_{\l 
\blQ}(t)|A^{\l\blQ}_{cv\blk}|^{2}
\label{xbecarr}
\ee
for conduction and valence states respectively.
For low and moderate excitation densities,
Eq.~(\ref{xbecarr}) 
highlights a striking departure from conventional Markovian
NEGF approaches, which produce carrier relaxation toward Fermi–Dirac
distributions. In contrast, the XBE framework predicts that the
occupations $N_{\l \mathbf{Q}}$ thermalize according to a Bose–Einstein
distribution evaluated at the e–h energies 
$E_{\l\mathbf{Q}}$~\cite{10.21468/SciPostPhys.18.1.009}.
Consequently, the resulting single-particle distributions
$f_{c \mathbf{k}}$ and $f_{v \mathbf{k}}$ do not converge to Fermi–Dirac forms;
instead, they adopt the momentum-space structure of exciton wavefunctions, signaling 
the formation of a correlated many-body state.

The XBE formalism builds on two main elements. First, the occupation
of the electron–hole eigenstates $N_{\l \blQ}$ is 
decomposed into coherent and incoherent 
components~\cite{PhysRevB.62.2706,brem,sangalli2018ab,10.21468/SciPostPhys.18.1.009},
\be
N_{\l \blQ}=\d_{\blQ ,
{\bf{0}}}N^{{\rm coh}}_{\l}+N^{{\rm inc}}_{\l \blQ}
\ee
which evolve according to separate dynamical equations.
The coherent part arises from the squared modulus of the excitonic polarization
$N^{{\rm coh}}_{\l}=|\r_{\l}|^{2}$, and is nonzero only for bright states
with zero center-of-mass momentum.
Second, determining both $\r_{\l}$ and $N^{{\rm inc}}_{\l \blQ}$
requires knowledge of the occupations of the
{\em irreducible} e-h states~\cite{exph4,10.21468/SciPostPhys.18.1.009} 
$\widetilde{N}_{\l \blQ}$.
These act as auxiliary variables within the XBE, 
preventing overscreening and ensuring 
a consistent treatment of many-body interactions.
The resulting XBE read~\cite{10.21468/SciPostPhys.18.1.009}
\bea
\frac{d}{dt}\r_{\l}&=&-iE_{\l \bz}\r_{\l}
-i\W_{\l}- \frac{1}{2}\sum_{\widetilde{\l}'\blQ}
\Big[\widetilde{\G}^{{\rm pol}}_{\l\widetilde{\l}'\blQ}
+\widetilde{\G}_{\l\widetilde{\l}'\blQ}
\widetilde{N}_{\widetilde{\l}'\blQ}
\Big]
\r_{\l} \nonumber \\
\frac{d}{dt}\widetilde{N}_{\widetilde{\l}\blQ}&=&
-\widetilde{\G}^{\rm out}_{\widetilde{\l}\blQ}\,
\widetilde{N}_{\widetilde{\l}\blQ}
+\widetilde{\G}^{\rm in}_{\widetilde{\l}\blQ}\,
\big(1+\widetilde{N}_{\widetilde{\l}\blQ}\big) \nonumber \\
\frac{d}{dt}N^{\rm inc}_{\l\blQ}&=&
-\G^{\rm out}_{\l\blQ}\,
N^{\rm inc}_{\l\blQ}
+\G^{\rm in}_{\l\blQ}\,
\big(1+N^{\rm inc}_{\l\blQ}\big)
+\sum_{\l'\widetilde{\l}}|S^{\blQ}_{\l\widetilde{\l}}|^{2}
\Big[\widetilde{\G}^{\rm pol}_{\l'\widetilde{\l}\blQ}
+\widetilde{\G}_{\l'\widetilde{\l}\blQ}\widetilde{N}_{\widetilde{\l}\blQ}
\Big]|\r_{\l'}|^{2}
\label{xbe}
\eea
The quantity $S^{\blQ}$ denotes the overlap between the 
wavefunctions  $A^{\l\blQ}$ obtained from Eq.~(\ref{eigeneqX}) and the 
wavefunctions of irreducible e-h states. The latter are computed by
solving a Bethe–Salpeter equation in which the HSEX kernel is 
replaced by the screened-exchange kernel~\cite{exph4}, 
see also Supporting Note~2.
Optical excitation enters through the driving term
$\W_{\l}=\sum_{cv\blk} A^{\l\bf{0}}_{cv\blk} \W_{cv\blk} $
expressed in terms of the Rabi frequencies $ \W_{cv\blk}$ associated with vertical
transitions from valence states $v\blk$ to conduction states $c\blk$.
Phonon-mediated scattering incorporates the effects of a sophisticated
self-energy that includes
T-matrix–type vertex corrections to the Fan–Migdal self-energy.
This advanced treatment yields a hierarchy of scattering rates that 
consistently capture multiple relaxation pathways.
Specifically, these rates account for the dephasing of the excitonic polarization 
($\widetilde{\G}^{{\rm pol}}$)
and inelastic transitions between e-h states.
The latter include both population loss via out-scattering processes  ($\G^{{\rm 
out}}$, $\widetilde{\G}^{{\rm out}}$) and population gain via 
in-scattering processes
($\G^{{\rm in}}$, $\widetilde{\G}^{{\rm in}}$, $\widetilde{\G}$).
Each rate is determined by the overlap of e-h 
wavefunctions weighted by the corresponding electron–phonon matrix elements, 
and all processes are constrained by energy conservation.
Detailed expressions for these rates are provided in the Supporting 
Note~2.
We notice that in the absence of photo-induced coherence
(i.e. for $\r_{\l}=0$), the XBE reduce to the excitonic Boltzmann 
equations, that describe the phonon-assisted dynamics of incoherent 
excitons only~\cite{PhysRevResearch.4.043203,PhysRevB.111.184305}
in terms of exciton-phonon 
couplings~\cite{Toyozawa,exph1,exph2,exph3,exph4}.

We assess the accuracy of  hot carrier dynamics
predicted by the XBE framework by directly
comparing the results with those obtained from SEPE. We recall that at low excitation 
density, the carrier dynamics described by SEPE is governed by 
conventional e-p Boltzmann scattering,
whereby relaxation toward the band extrema proceeds predominantly via 
phonon emission.
In this work we approximate the quasi-free 
continuum eigenstates of Eq.~(\ref{eigeneqX})
as noninteracting e-h  pairs, with energies
 $E_{\l\blQ} \approx \e_{c_{\l}\blk_{\l}+\blQ}-\e_{v_{\l} 
\blk_{\l}}$, and wavefunctions
$A^{\l\blQ}_{cv\blk} \approx 
\d_{c,c_{\l}}\d_{v,v_{\l}}\d_{\blk,\blk_{\l}} $,
where the triplet $(v_{\l},c_{\l},\blk_{\l})$ is uniquely specified by the branch index $\l$.
Within this approximation, intra-continuum relaxation rigorously reduces 
to conventional Fan–Migdal electron–phonon scattering - the microscopic
mechanism underpinning the SEPE framework (see Supporting Note 3). 
This establishes a direct and nontrivial equivalence between XBE 
and the standard theory of phonon-mediated carrier 
relaxation in the dilute limit, representing a 
central result of this work.
Accordingly, if the bound solutions of Eq.~(\ref{eigeneqX})
are neglected, 
the carrier occupations given by Eq.~(\ref{xbecarr}) recover those of 
SEPE.
Crucially, however, the XBE incorporate additional scattering 
channels 
that capture the conversion from free to bound electrons,  driving 
the system toward occupation
of the lowest-energy excitonic levels.
Hereinafter, we use the term {\em bound electrons} 
to refer specifically to electrons in bound excitonic states.

\subsection{Carrier dynamics following sudden excitation}

Monolayer WSe$_{2}$ provides an ideal platform for investigating excitonic 
effects in ultrafast carrier dynamics. In this direct-gap material,
the global minimum of the conduction band is located at the K and K'
points of the first Brillouin zone. However, six additional local minima 
occur at the Q points, separated in energy by only a few tens of 
meV~\cite{wse2gw} (see Fig.~\ref{fig1}b).
This near-degeneracy gives rise to a rich manifold of tightly bound excitonic
states.
Because the conduction-band extrema  at K and Q are energetically close, 
excitons can form with center-of-mass momenta near $\Gamma$, K, Q, and M,
resulting in a series of quasi-degenerate intra- and
inter-valley excitons~\cite{Deilmann_2019} (see also Fig.~\ref{fig1}c).
These excitonic states predominantly involve holes
residing in the K valleys, while the corresponding electrons occupy 
either the K or Q valleys.
As we demonstrate below, the finite energy splitting between the 
single-particle conduction-band minima at K and Q, contrasted with 
the near-degeneracy of the corresponding excitons,
plays a decisive role in shaping different carrier dynamics in SEPE 
and XBE.

We first investigate the carrier dynamics triggered 
by the instantaneous generation of hot electron and 
hole populations, at a lattice temperature $T=70~$K (see Supporting 
Note~4 for details about the simulations).
The initial state is characterized
by a small excitation density 
$n=\frac{1}{N_{\blk}A_{{\rm cell}}}\sum_{c\blk}f_{c\blk}=10^{11}{\rm cm}^{-2}$
(here $N_{\blk}$ is the number of $\blk$-points and $A_{{\rm 
cell}}=9.6~\AA^{2}$ 
is the area of the unit cell of WSe$_{2}$)
and is modeled by a narrow Gaussian distribution
centered at an initial e-h excitation energy of 2.4~eV 
(the bandgap being approximately 2.05~eV~\cite{madeo2020,wse2gw}). 
This initial condition corresponds to the standard setup
employed in BE–based simulations, where the
dynamics are initiated from incoherent carrier populations
with excess kinetic energy, and the
pump-induced interband coherence is explicitly neglected.
The subsequent evolution reflects purely
phonon-driven relaxation processes, disentangled from coherent 
light–matter coupling effects. Given the low excitation density, 
the energy transferred to the lattice is negligible, and 
it is therefore an excellent approximation to treat the lattice temperature
as constant during carrier relaxation~\cite{perfetto_2023}.

   \begin{figure}[tbp]
   \centering
   \includegraphics[width=0.99\textwidth]{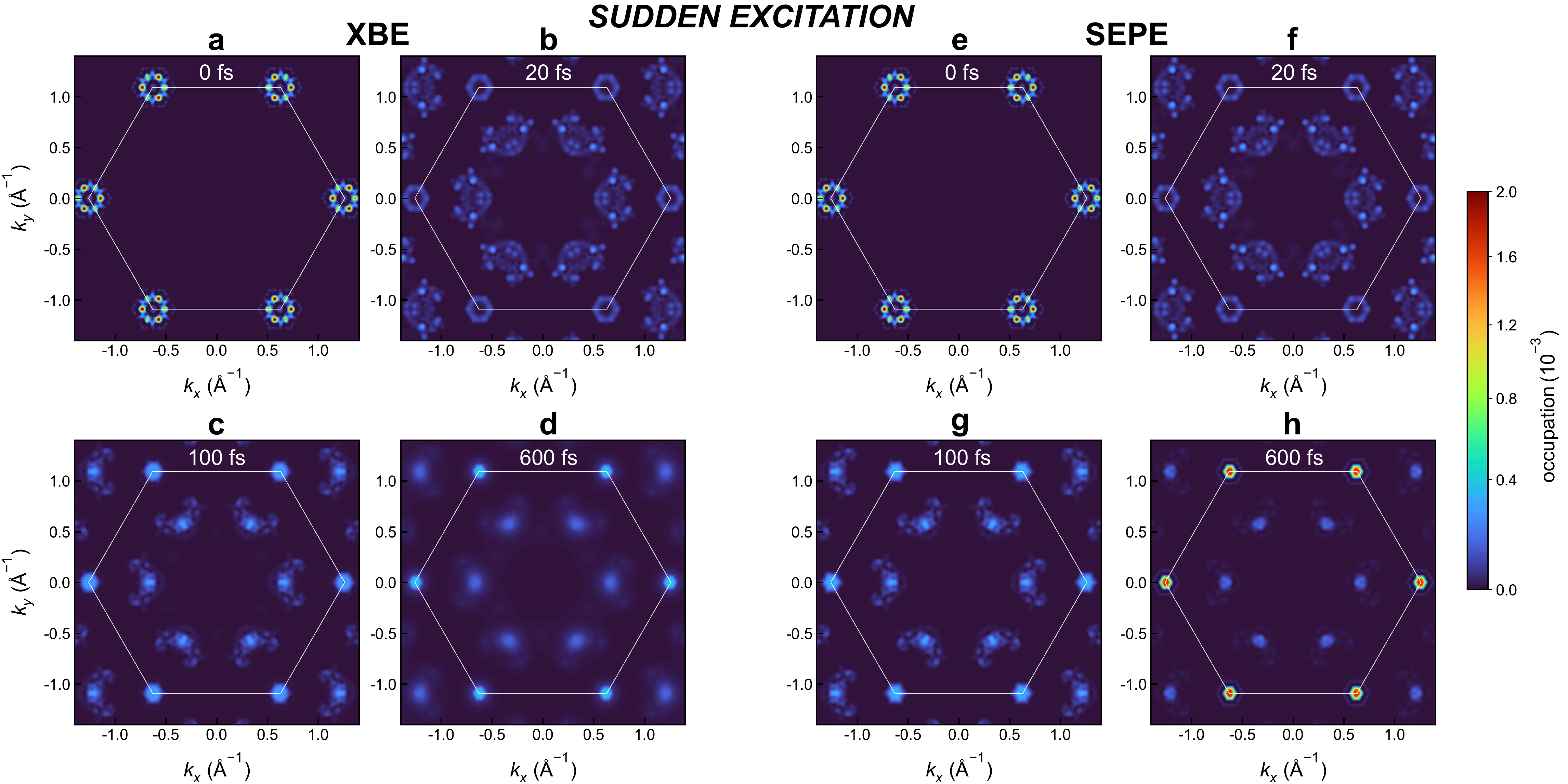}
   \caption{Density plot of the momentum-resolved conduction-band 
   occupation, $f_{C\mathbf{k}}$,
   at selected times (0 fs, 20 fs, 100 fs, and 600 fs) following the 
   sudden excitation described in
   the main text. Results are shown for the XBE framework (panels a–d) and 
   the SEPE framework (panels e–h).
   }
   \label{fig2}
   \end{figure}

\begin{figure}[tbp]
   \centering
   \includegraphics[width=0.99\textwidth]{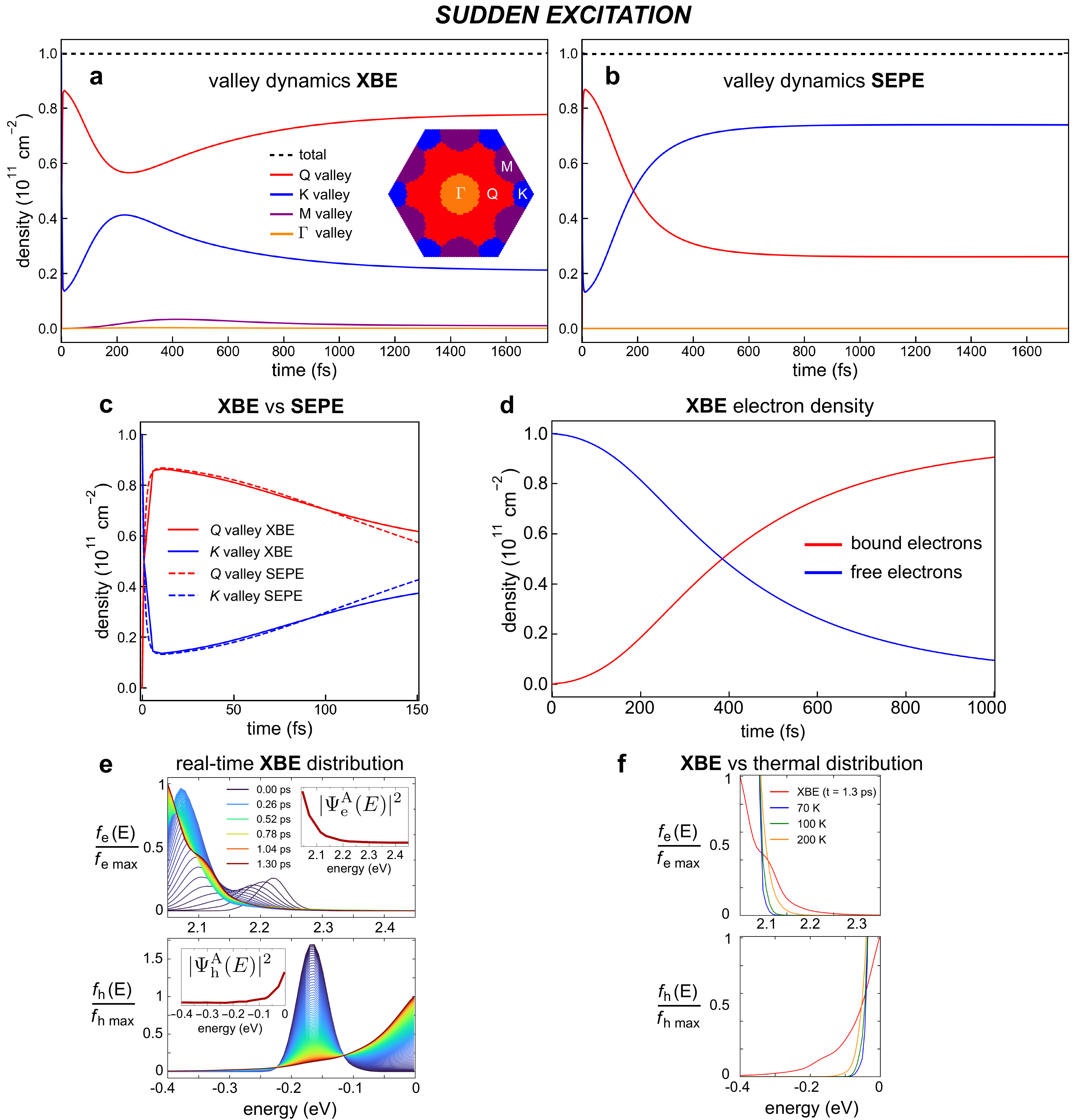}
   \caption{Carrier dynamics following the sudden excitation.
   Panel a: Time-dependent carrier populations in the K, Q, M, and
   $\Gamma$ valleys of monolayer WSe$_{2}$ 
   (indicated in the inset), calculated within the XBE framework.
Panel b: Same as panel a, but obtained within the SEPE framework.
Panel c: Comparison of the early-time valley dynamics predicted by
the XBE and SEPE approaches.
Valley populations are evaluated as described in the Supporting 
Note~4.
Panel d: Time-dependent conduction  populations 
   corresponding to free $n_{{\rm f}}$ (blue) and bound $n_{{\rm b}}$ 
   electrons 
   (red).
Panel e: Energy-resolved conduction-electron and valence-hole 
occupations, $f_{\mathrm{e}}(E)$ 
(top) and $f_{\mathrm{h}}(E)$ (bottom), at selected times in the range
$0$–$1.3$ ps. The occupations $ f_{\mathrm{e/h}}(E)$ are averaged over 
spin and normalized 
with respect the corresponding maximum values $ 
f_{\mathrm{e/h\,max}}(E)$ reached at time $t=1.3~$ps.
In the insets we show the envelope $|\Psi^{{\rm A}}_{{\rm e/h}}(E)|^{2}$
of the square modulus of the A-exciton wavefunction as a function of 
conduction (top) and valence (bottom) band-energy.   
Panel f: Comparison of $f_{\mathrm{e}}(E)$ and $f_{\mathrm{h}}(E)$ 
at $1.3$ ps with Fermi–Dirac distributions at temperatures $T = 70$, $100$, and $200$ K.
   }
   \label{fig3}
   \end{figure}

In Fig.~\ref{fig2} we illustrate the rich multivalley dynamics
of the conduction-band population $f_{C\blk}=\sum_{c}f_{c\blk}$
across the first Brilloiun zone.
The momentum-resolved hole dynamics in the valence bands (not shown) is 
comparatively simpler, as it involves only the K valley.
At time $t=0$ 
(Fig.~\ref{fig2}a), the hot carriers form a ring-shaped distribution 
around the K and K' points, 
reflecting their excess kinetic energy.
The early transient dynamics is dominated by  
intervalley scattering~\cite{wallauer2016,madeo2020,dong2021direct,Wallauer2},
which rapidly redistributes carriers into the six
Q valleys that have comparable energy.
This process is mediated by zone-edge optical phonons~\cite{waldecker2017}
and is known to be highly
efficient in transition-metal dichalcogenides.
Already within $20~$fs a population inversion between K and Q valleys is established,
with nearly 90\% of the hot carriers transferred to Q, see 
Fig.~\ref{fig3}a,b. 
Following this ultrafast intervalley transfer, carriers gradually migrate 
back towards K on a slower timescale.  In this regime, intervalley scattering 
competes with intravalley relaxation,
which drives electrons towards the local band minima via 
the emission of small-momentum phonons.
By $t\approx 100~$fs pronounced population maxima have developed at the Q points,
whereas accumulation at K remains comparatively slow, see 
Fig.~\ref{fig2}c,g.
At this stage, the carriers are still well described as 
a quasi-free electron-hole plasma, and the dynamics predicted by
the XBE and SEPE approaches are essentially indistinguishable 
(Fig~\ref{fig3}c).
For  $t \gtrsim 150~$fs, however, the progressive accumulation of carriers 
near the band extrema  promotes efficient formation 
of bound electrons.
The temporal evolution of the electron density obtained 
from the XBE is shown in Fig.~\ref{fig3}d, separately
resolving contributions from free electrons and electrons bound to 
holes. The density of free and bound electrons is obtained as 
$n_{{\rm f/b}}=\frac{1}{N_{\blk}A_{{\rm cell}}}\sum_{c\blk}f_{c\blk}$, where 
in calculating $f_{c\blk}$ from Eq.~(\ref{xbecarr}), the sum over $\l$
is restricted to continuum and bound  states respectively.
As shown in panel d, by $t = 200~$fs bound electrons account for approximately
20\% of the total population.
In this regime, excitonic correlations substantially reshape the 
carrier relaxation
pathways, and the predictions of the two approaches begin to diverge.
Within the XBE framework, the dynamics becomes increasingly governed by scattering rates 
$\Gamma^{{\rm in/out}}$ involving  bound electrons (see 
also Supporting Note~2)
leading to two distinct effects.

First, the back-scattering towards the K valley is markedly slowed down.
Excitons with the electron localized near K and Q are nearly degenerate
in energy; as a result, the long-time populations of the individual K and Q 
valleys become comparable. However, the larger valley multiplicity of Q 
(sixfold) relative to K (twofold) amplifies the total occupation of Q,
which ultimately exceeds that of K by approximately a factor of three.
Transient carrier
redistribution  is therefore governed by excitonic phase-space constraints.
The build-up of valley population imbalance is reflected in a non-monotonic population
flow: around $\sim 200~\mathrm{fs}$, the Q population reaches
a minimum while the K 
population exhibits a corresponding maximum (see Fig.~\ref{fig3}a).
By contrast, the SEPE dynamics is controlled by the single-particle
band structure. Since the absolute conduction-band minimum resides
at K, whereas the Q valleys lie higher in energy by approximately
$40~\mathrm{meV}$, Fan–Migdal electron–phonon scattering preferentially
drives carriers toward K. This results in a monotonic population
dynamics between Q and K and, consequently, an opposite valley 
polarization (see Fig.~\ref{fig3}b). This underscores the fundamentally
different relaxation mechanisms captured by the two approaches.

The second key effect concerns intravalley thermalization 
and the nature of the emerging quasi-equilibrium state.
After approximately 1~ps, the momentum-resolved occupations become
quasi-stationary, evolving  slowly toward their asymptotic values.
Within the SEPE framework, we have previously demonstrated~\cite{sepe} that the
long-time carrier distributions converge to Fermi–Dirac
statistics at the lattice temperature, $T = 70~\mathrm{K}$, 
set by the phonon bath.
In stark contrast, the XBE dynamics yields a qualitatively different 
steady state. Thermalization proceeds according to Eq.~(\ref{xbecarr}),
where the relevant degrees of freedom are e-h occupations $N_{\l \mathbf{Q}}$
that obey Bose–Einstein statistics at $T = 70~\mathrm{K}$.
Crucially, in this correlated regime electrons and holes no longer behave as 
independent quasiparticles, but persist as the fermionic constituents of bound excitons.
Figure~\ref{fig3}e shows the time evolution of the energy-resolved
electron and hole distributions, 
$f_{\mathrm{e}}(E)=\nu^{-1}(E)\sum_{c\blk:\e_{c\blk}=E}f_{c\blk}$ and
$f_{\mathrm{h}}(E)=\nu^{-1}(E)\sum_{v\blk:\e_{v\blk}=E}f_{v\blk}$
for conduction and valence bands respectively [$\nu(E)$ being the density of states].
A prominent feature emerging at long
times is the development of a kink in $f_{\mathrm{e}}(E)$ 
at $E \sim 2.1~\mathrm{eV}$. This 
reflects a substantial steady-state population at the bottom
of the Q valleys, despite their higher energy relative to the 
K valleys. 
Such behavior originates from the comparable populations of 
quasi-degenerate excitons,
in which the electron resides in either the K or Q valleys.
This excitonic character also manifests in the shape of the single-particle
momentum distributions, which inherit the structure of the underlying
exciton wavefunctions. Remarkably, the asymptotic forms of
$f_{\mathrm{e}}(E)$ and $f_{\mathrm{h}}(E)$ closely resemble
the squared modulus of the A-exciton wavefunction 
in conduction band energy space  
$|\Psi^{{\rm A}}_{{\rm 
e}}(E)|^{2}=\nu^{-1}(E)\sum_{c\blk:\e_{c\blk}=E}\sum_{v}|A^{{\rm A}}_{cv 
\blk}|^{2}$ (with $A^{{\rm A}}_{cv \blk}$ the A-exciton wavefunction)  
and valence-band energy space $|\Psi^{{\rm A}}_{{\rm 
h}}(E)|^{2}=\nu^{-1}(E)\sum_{v\blk:\e_{v\blk}=E}\sum_{c}|A^{{\rm A}}_{cv 
\blk}|^{2}$ (see insets of Fig.~\ref{fig3}e). 
Because tightly bound excitons are intrinsically delocalized in momentum 
space, their wavefunction envelopes span a broad range of $\mathbf{k}$
values~\cite{man2021experimental,https://doi.org/10.1002/ntls.10010}. 
This directly translates into markedly
broadened carrier distributions with pronounced high-energy tails, 
even at low temperatures.
A direct comparison with Fermi–Dirac distributions (Fig.~\ref{fig3}f) 
highlights the correlated nature of this steady state. The long-time 
carrier distributions obtained within the XBE framework cannot be 
captured by any description based on non-interacting thermal fermions,
even when invoking artificially elevated temperatures.

 \begin{figure}[tbp]
   \centering
   \includegraphics[width=0.99\textwidth]{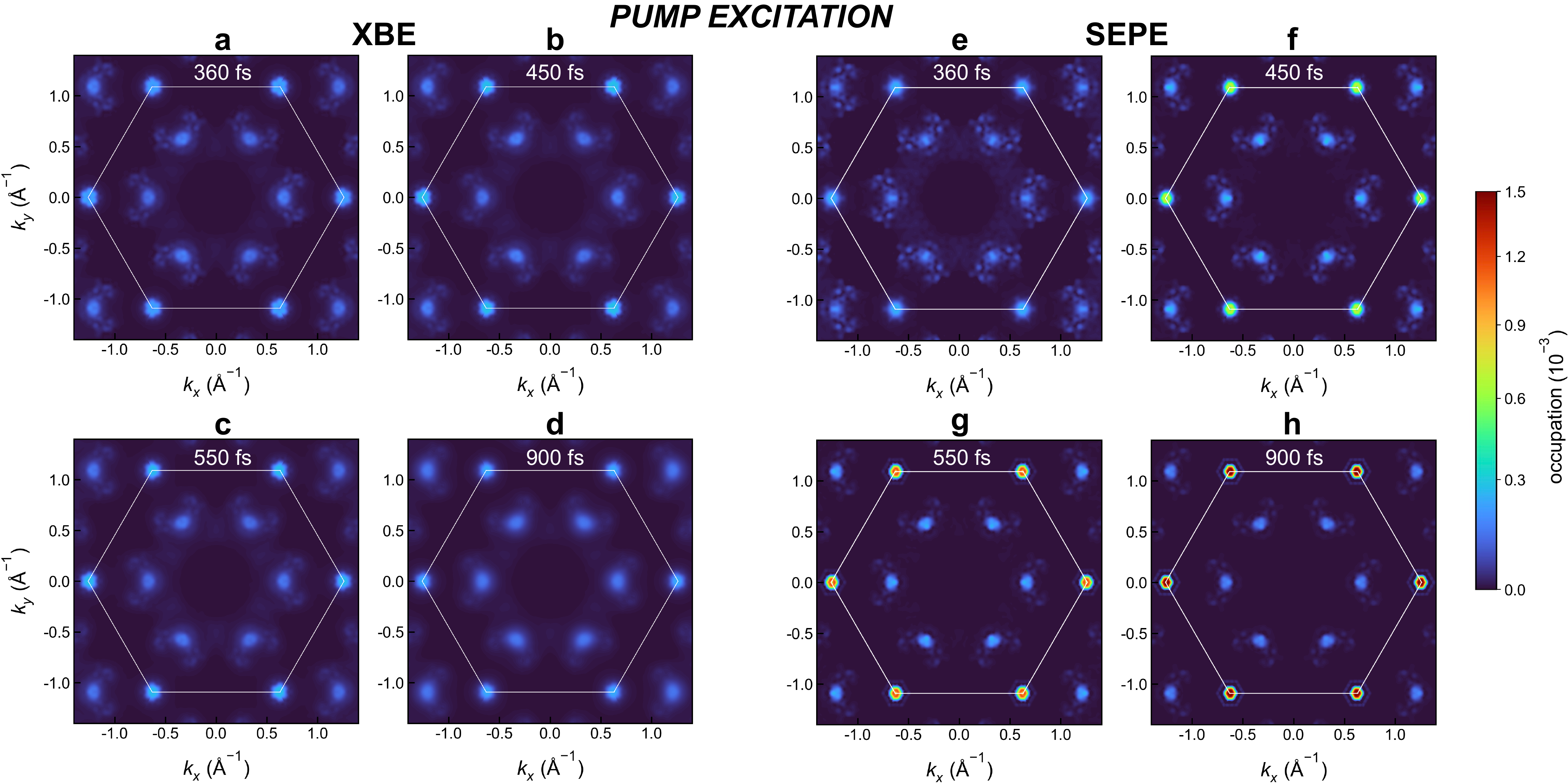}
   \caption{Density plot of the momentum-resolved conduction-band 
   occupation, $f_{C\mathbf{k}}$,
   at selected times (360 fs, 450 fs, 550 fs, and 900 fs) following the 
   pump excitation described in
   the main text. Results are shown for the XBE framework (panels a–d) and 
   the SEPE framework (panels e–h).
   }
   \label{fig5}
   \end{figure}

\subsection{Pump-induced carrier dynamics}
   
We now turn to the experimentally relevant scenario 
in which photoexcitation is driven by an external laser field.
Within the XBE framework, the interplay between light-induced 
coherence and e-p interaction is intrinsically nontrivial.
According to Eq.~(\ref{xbe}), the pump pulse first generates 
e-h polarizations $\rho_{\l}$. 
These coherent amplitudes subsequently decay due to e–p scattering, 
and their decay is accompanied by a corresponding buildup
of incoherent e-h populations~\cite{PhysRevB.62.2706,brem,10.21468/SciPostPhys.18.1.009}.
Crucially, this polarization-to-population
transfer is mediated by the coupling to irreducible e-h states 
(see $\widetilde{\G}^{{\rm pol}}$  rates),
a mechanism that emerges when the screening of both e-e and 
e-p interactions is treated consistently.
Within this conserving description, the total number
of photoexcited e-h pairs-- counting both coherent and incoherent 
contributions -- is preserved once the pump field has vanished.

   \begin{figure}[tbp]
   \centering
   \includegraphics[width=0.99\textwidth]{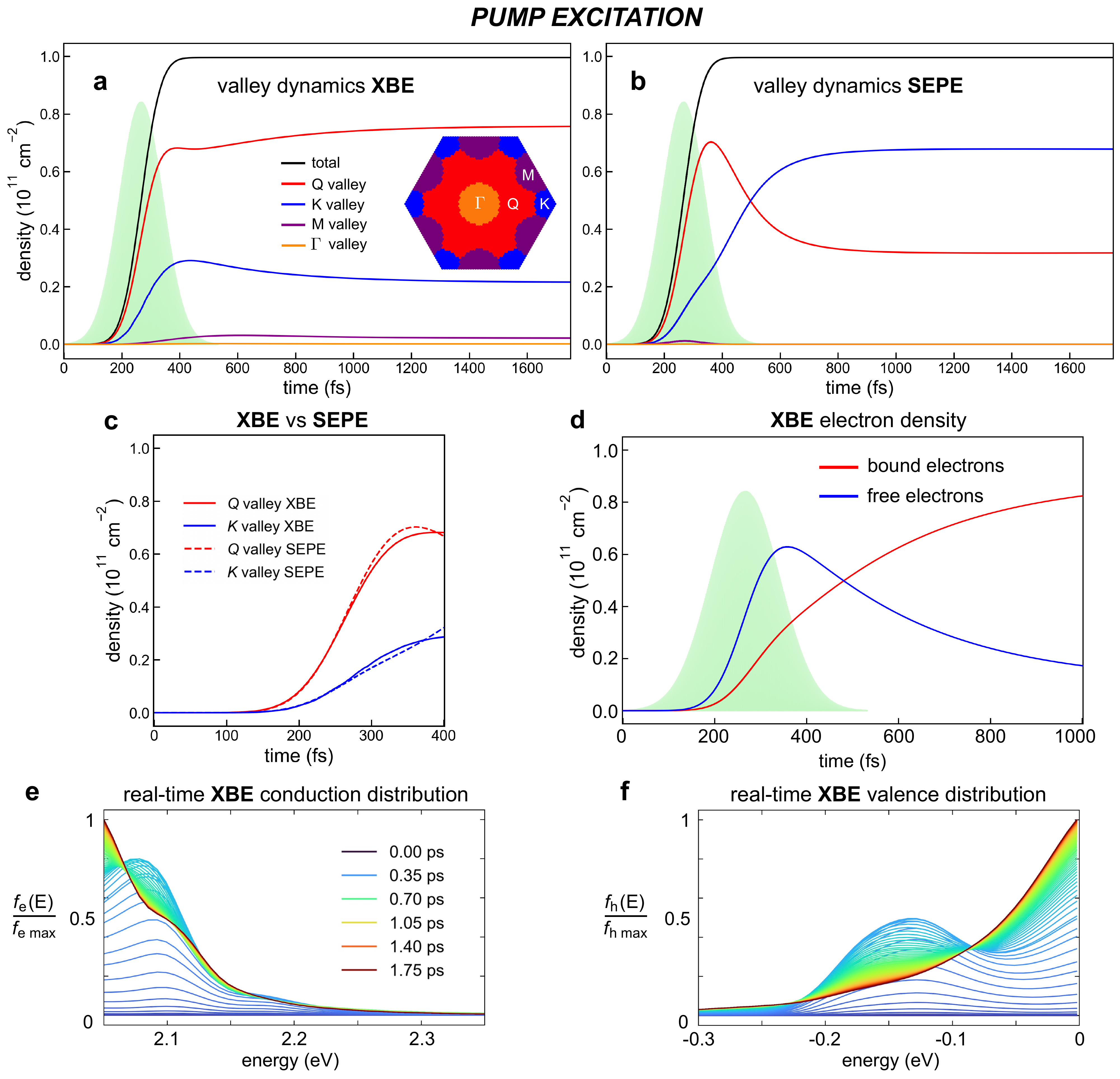}
   \caption{Carrier dynamics following the pump excitation.
   Panel a: Time-dependent carrier populations in the K, Q, M, and
   $\Gamma$ valleys of monolayer WSe$_{2}$ 
   (indicated in the inset), calculated within the XBE framework. The 
   temporal profile of the pump pulse is shown as a green shaded 
   region.
Panel b: Same as panel a, but obtained within the SEPE framework.
Panel c: Comparison of the early-time valley dynamics predicted by
the XBE and SEPE approaches.
Valley populations are evaluated as described in the Supporting 
Note~4.
Panel d: Time-dependent conduction  populations 
   corresponding to free $n_{{\rm f}}$ (blue) and bound $n_{{\rm b}}$ 
   electrons 
   (red).
Panel e: Energy-resolved electron and hole occupations, $f_{\mathrm{e}}(E)$ 
(top) and $f_{\mathrm{h}}(E)$ (bottom), at selected times in the range
$0$–$1.75$ ps. The occupations $ f_{\mathrm{e/h}}(E)$ are  averaged over 
spin and normalized 
with respect the corresponding maximum values $ 
f_{\mathrm{e/h\,max}}(E)$ reached at time $t=1.75~$ps.
   }
   \label{fig6}
   \end{figure}
 
Motivated 
by the experiment of Ref.~\cite{madeo2020}, we consider a linearly polarized 
pump pulse with central photon energy $2.4~\mathrm{eV}$ and duration
$250~\mathrm{fs}$, exciting free carriers well above the band gap. 
The pulse fluence is adjusted to generate an excitation density 
of $n = 10^{11}~\mathrm{cm^{-2}}$,
matching the conditions of the previous analysis (see Supporting 
Note~4  for details about the simulations).
Figure~\ref{fig5} shows selected temporal snapshots of the momentum-resolved
conduction-band population $f_{C\mathbf{k}}$. Remarkably, a highly
nontrivial redistribution of carriers takes place already during
the pump action. Efficient intervalley
scattering transfers electrons from the K valleys to the energetically
nearby Q valleys on ultrafast timescales, i.e., concomitantly with
their photogeneration. At $t \sim 250~\mathrm{fs}$, when the pulse
reaches its maximum amplitude, the total population in Q is 
already approximately twice that in K, see Fig.~\ref{fig6}a,b. In this early transient
regime, e-h correlations have
not yet fully developed. Accordingly, the XBE and SEPE approaches
yield very similar results, indicating that the dynamics 
is still dominated by single-particle scattering processes rather 
than by excitonic effects (Fig.~\ref{fig6}c).
Consistently, Fig.~\ref{fig5}d shows that at this time the population of bound 
electrons is approximately
30\% of the total population.
Nevertheless, in this temporal window the bound-electron formation rate is maximal,
as rapid intra-valley scattering drives
electrons toward the band extrema, thus 
enhancing correlations.
As a result, for $t \gtrsim 250~\mathrm{fs}$ the two theoretical 
descriptions begin to diverge.
Within XBE, the progressive buildup of excitonic correlations opens
correlated scattering pathways that further enhance
the occupation of the Q valleys, reinforcing the
imbalance established during the pump (Fig.~\ref{fig6}a).
The SEPE approach, intead, predicts the opposite trend: 
constrained by the single-particle energy landscape,
electrons in the Q valley start to migrate back to the K valleys, 
thereby inverting the transient valley polarization (Fig.~\ref{fig6}b).
Strikingly, by the end of the pump pulse ($t \approx 500~\mathrm{fs}$), 
the two scenarios have already separated qualitatively. The XBE 
approach yields a Q-valley
population approximately three times larger
than that of K, whereas SEPE predicts that the K valleys host roughly twice
as many electrons as the Q valleys.
Importantly, the XBE prediction is in quantitative agreement with the time-resolved
ARPES measurements reported in Ref.~\cite{madeo2020}.
Specifically, after post-processing the ARPES spectra to extract the transient carrier populations
from the experimental spectra yields a K-to-Q ratio of $\sim 0.3$,
strikingly consistent with our XBE results.
This experimental consistency provides strong support for a 
description in which excitonic 
correlations, already during the pump action, decisively reshape 
the valley-resolved carrier dynamics.
In the final stage of the dynamics, the system gradually thermalizes
toward quasi-equilibrium, following the same
qualitative pathway identified for the sudden-excitation scenario 
(Fig.~\ref{fig6}e,f).

\section{Conclusions}

We have developed a two-particle Green's-function framework
that explicitly incorporates excitonic correlations into the 
nonequilibrium photocarrier dynamics of semiconductors, overcoming
a central limitation of conventional single-particle approaches. 
Built upon the recently formulated excitonic Bloch 
equations~\cite{10.21468/SciPostPhys.18.1.009}, the
method provides a controlled and quantitatively reliable description
of photoexcited systems in the low-excitation regime. 
Electron–hole interactions are treated in a nonperturbative fashion,
within a  
many-body scheme that includes the effects of T-matrix–type 
vertex corrections to the Fan–Migdal self-energy.
Exciton–phonon scattering processes are naturally incorporated, 
allowing the carrier dynamics to be governed by the rich 
electron–hole energy landscape rather than being constrained by
the underlying single-particle band structure.
A central result of this work is the direct comparison with state-of-the-art
semiconductor electron–phonon equations~\cite{sepe}.
We demonstrate that, when bound-state formation is suppressed, the
excitonic Bloch equations rigorously reproduce the conventional electron–phonon 
scattering as formulated in the Boltzmann equations. 
Consequently, XBE and SEPE yield identical predictions as long
as carrier dynamics are governed by 
phonon-mediated cooling of quasi-free electrons and holes.
However, when applying the method to non-resonantly excited monolayer WSe$_{2}$, 
our numerical results show that this agreement is 
limited to the initial stage of carrier relaxation. As
carriers accumulate near the band extrema e-h correlations
open efficient relaxation channels toward bound states.
When bound electrons account for roughly 30\% of the total population,
the predictions of the excitonic Bloch equations and SEPE begin to diverge markedly.
Remarkably, this stage is reached already during the action of a typical pump pulse with 
a duration of a few hundred femtoseconds.
In this regime, the excitonic
Bloch equations predict valley populations
consistent with time-resolved ARPES measurements~\cite{madeo2020}, whereas conventional 
approaches yield qualitatively different trends.
More fundamentally, our results demonstrate that the long-time state of
a photoexcited excitonic semiconductor cannot be interpreted as 
a thermalized Fermi gas of independent electrons and holes. Instead, the
system evolves toward a correlated quasi-equilibrium dominated by excitonic
populations. 
Here carrier momentum-distributions inherit the structure
of excitonic wavefunctions rather than Fermi–Dirac statistics.
Notice that the lowest-energy excitons  dominating
the thermalized population are not necessarily optically bright. 
Indeed, in the specific case considered here, the lowest-energy excitons 
are all dark.
The  emergence of this quasi-equilibrium state is a robust, general
phenomenon  in  excitonic materials, and is largely independent 
of the excitation protocol. 
Consistently, our simulations show that the final carrier
distributions are essentially identical whether the system is
initialized through a sudden excitation or through an explicit
pump-driven process.
Our
findings thus  reshape the conventional picture of thermalization in photoexcited
semiconductors and establishes electron–hole correlations as a governing 
factor well beyond the ultrafast formation regime.
Ultimately, the present framework provides a predictive microscopic 
route for describing relaxation and thermalization in excitonic 
materials, bridging ultrafast many-body theory and experiment.

\section*{Acknowledgements}

The authors  acknowledge funding from Ministero Università e Ricerca PRIN 
under grant agreement No. 2022WZ8LME, from INFN through project TIME2QUEST, 
from the
Horizon Europe research and innovation program of
the European Union under the Marie Sk\polishl{}odowska-Curie
grant agreement 101118915 (TIMES),
from 
Tor Vergata University through project TESLA, and from CINECA through the IsCd3 project EEPCDTDS.
The authors  also thank C. Attaccalite and D. Santos Stone for 
fruitful discussions and B. Demoulin and  A. Saul for the management of the computer cluster \emph{Claudia}.

% 
% Please use ``The authors thank \ldots'' rather than ``The authors would like to
% thank \ldots''.

\section*{Supporting information}

% A listing of the contents of each file supplied as Supporting Information
% should be included. For instructions on what should be included in the
% Supporting Information as well as how to prepare this material for
% publications, refer to the journal's Instructions for Authors.
% 
% The following files are available free of charge.
% \begin{itemize}
%   \item Filename-1: brief description
%   \item Filename-2: brief description
% \end{itemize}

The Supporting information contains:
 \begin{itemize}
   \item Supporting Note~1: Carrier populations from 
excitonic Bloch equations.
   \item Supporting Note~2: Excitonic scattering rates
   \item Supporting Note~3: Recovering the semiconductor Bloch equations 
   from the excitonic Bloch equations
   \item Supporting Note~4: First-principles simulations
 \end{itemize}

%%%%%%%%%%%%%%%%%%%%%%%%%%%%%%%%%%%%%%%%%%%%%%%%%%%%%%%%%%%%%%%%%%%%%
%% If you are using classical BibTeX rather than biblatex,
%% remove the \printbibliography and uncomment the \bibliograpy one
%%%%%%%%%%%%%%%%%%%%%%%%%%%%%%%%%%%%%%%%%%%%%%%%%%%%%%%%%%%%%%%%%%%%%
%\printbibliography
%\bibliography{acs-template.bib}
%\bibliography{mybiblio_new.bib}

%\addbibresource{mybiblio_new.bib}

% \newpage
% 
% \rule{0.05in}{1.75in}%
% \begin{minipage}[b][1.75in]{3.25in}
%   \sffamily
%   \frenchspacing
% 
%   Some journals require a graphical entry for the Table of Contents. This
%   should be laid out ``print ready'' so that the sizing of the text is correct.
% 
%   The space available depends on the journal: J. Am. Chem. Soc. allows 3.25 in
%   by 1.75 in and requires sanserif text. Some journals want different sizes:
%   you can easily adjust here.
%   
%   The two rules either side of the content are there to help judge the height
%   of your material: they may be deleted once not required.
%   
% \end{minipage}%
% \rule{0.05in}{1.75in}

\end{document}